\pdfoutput=1

\documentclass[11pt]{article}

\usepackage{acl}

\usepackage{times}
\usepackage{latexsym}

\usepackage[T1]{fontenc}

\usepackage[utf8]{inputenc}

\usepackage{microtype}

\usepackage{inconsolata}

\usepackage{graphicx}
\usepackage{hyperref}
\usepackage{xcolor}
\usepackage{amsmath}  
\usepackage{wrapfig,lipsum,booktabs}
\usepackage{multirow} 
\usepackage{wrapfig}
\usepackage[ruled,norelsize]{algorithm}
\usepackage{algpseudocode,algorithmicx}
\usepackage{textcomp}
\usepackage{framed}
\usepackage{todonotes}
\usepackage{booktabs}
\usepackage{tabularx} 
\usepackage{arydshln} 
\definecolor{darkblue}{rgb}{0,0,0.6}
\newcommand{\blt}[1]{\textcolor{black}{#1}}
\newcommand{\bluetexttt}[1]{\textcolor{black}{\texttt{#1}}}
\newcommand{\relup}[1]{\textcolor{blue}{+#1\%}}

\newcommand{\NameQR}{GenQREnsemble}
\newcommand{\NameDF}{GenQRFusion}
\newcommand{\NamePRF}{GenQREnsemble-RF}
\newcommand{\NameDFPRF}{GenQRFusion-RF}

\title{Generative Query Reformulation Using Ensemble Prompting, Document Fusion, and Relevance Feedback}

\author{Kaustubh D. Dhole, Ramraj Chandradevan, Eugene Agichtein \\
        Department of Computer Science \\ Emory University \\ Atlanta, USA-30307\\
        \texttt{\{kaustubh.dhole,rchan31,eugene.agichtein\}@emory.edu}}

\begin{document}
\maketitle
\begin{abstract}
Query Reformulation (QR) is a set of techniques used to transform a user’s original search query to a text that better aligns with the user’s intent and improves their search experience. Recently, zero-shot QR has been a promising approach due to its ability to exploit knowledge inherent in large language models. Inspired by the success of ensemble prompting strategies which have benefited other tasks, we investigate if they can improve query reformulation. In this context, we propose two ensemble-based prompting techniques, \textbf{\NameQR{}}\footnote{The extended work of~\citet{dhole2024genqrensemble}, European Conference on Information Retrieval, 2024} and \textbf{\NameDF{}} which leverage paraphrases of a zero-shot instruction to generate multiple sets of keywords to improve retrieval performance ultimately. We further introduce their post-retrieval variants to incorporate relevance feedback from a variety of sources, including an oracle simulating a human user and a ``critic'' LLM. We demonstrate that an ensemble of query reformulations can improve retrieval effectiveness by up to 18\% on nDCG@10 in pre-retrieval settings and 9\% on post-retrieval settings on multiple benchmarks, outperforming all previously reported SOTA results. We perform subsequent analyses to investigate the effects of feedback documents, incorporate domain-specific instructions, filter reformulations, and generate fluent reformulations that might be more beneficial to human searchers. Together, the techniques and the results presented in this paper establish a new state of the art in automated query reformulation for retrieval and suggest promising directions for future research. 
\end{abstract}

\section{Introduction}
Users searching for relevant documents might not always be able to express their information needs in their initial queries accurately. This could result in queries being vague or ambiguous or lacking the necessary domain vocabulary. Query Reformulation (QR) is a set of techniques used to transform a user’s original search query to a text that better aligns with the user’s intent and improves their search experience. Such reformulation alleviates the vocabulary mismatch problem by expanding the query with related terms or paraphrasing it into a suitable form by incorporating additional context.

Recently, with the success of large language models (LLMs)~\cite{gpt3,i2}, a plethora of QR approaches have been developed. The generative capabilities of LLMs have been exploited to produce novel queries~\cite{doc2query}, as well as useful keywords to be appended to the users' original queries~\cite{zeroshot}. By gaining exposure to enormous amounts of text during pre-training, prompting has become a promising avenue for utilizing knowledge inherent in an LLM for the benefit of subsequent downstream tasks~\cite{bigbench} especially QR~\cite{promptingqueryexpa,whenfail}. 

Unlike training or few-shot learning, zero-shot prompting does not rely on any labeled examples. The advantage of a zero-shot approach is the ease with which a standalone generative model can be used to reformulate queries by prompting a templated piece of instruction along with the original query. Particularly, zero-shot QR can be used to generate keywords by prompting the user's original query along with an instruction that defines the task of query reformulation in natural language like~\bluetexttt{Generate useful search terms for the given query:`List all the pizzerias in New York'}.

\begin{table*}
  \centering
  \resizebox{\textwidth}{!}{ %
    \begin{tabularx}{\textwidth}{XX}
         \textbf{Instruction} & \textbf{Generations}\\ \hline
         Increase the search efficacy by offering 
         beneficial expansion keywords for the query & age  goldfish  grow  outsmart  outlive ageing  species goldfish  grows  diet ...\\ \hline
         Enhance search outcomes by recommending beneficial expansion terms to supplement the query & Goldfish breed sizes What kind of goldfish grows the fastest Do goldfish have scales ...\\ \hline
         Optimize search results by suggesting meaningful expansion terms to enhance the query & Goldfish genus  Betta bonsai or Fancy goldfish also known as Loachyodidae Family... \\ 
    \end{tabularx}
    }
    \caption{Reformulations generated for the query (``do goldfish grow'') differ drastically when generated from three paraphrastic instructions prompted to~\bluetexttt{flan-t5-xxl}.}
    \label{tab:example}
\end{table*}

However, such a zero-shot prompting approach is still contingent on the exact instruction appearing in the prompt providing plenty of avenues of improvement. While LLMs have been known to vary significantly in performance across different prompts~\cite{robust1,robust2} and generation settings~\cite{decoding}, many natural language tasks have benefited by exploiting such variation via ensembling multiple prompts or generating diverse reasoning paths~\cite{diverse,ama,selfcons}. Whether such improvements also transfer to tasks like QR is yet to be determined. We hypothesize that QR might naturally benefit from prompt variation -- An ensemble of zero-shot reformulators with paraphrastic instructions can be tasked to look at the input query in diverse ways to elicit different expansions. This work proposes the following contributions:
\begin{itemize}
    \item We propose two novel pre-retrieval methods, \textbf{\NameQR{}} and \textbf{\NameDF{}} -- zero-shot \textbf{Ensemble} based \textbf{Gen}erative  \textbf{Q}uery \textbf{R}eformulator and \textbf{Doc}ument \textbf{F}usion  -- which exploit multiple zero-shot instructions for QR generating more effective query reformulations than possible with a single instruction.
    \item We further introduce their post-retrieval extensions \textbf{\NamePRF{}} and \textbf{\NameDFPRF{}} to incorporate \textbf{R}elevance \textbf{F}eedback from a variety of sources, including human searchers and ``critic'' LLMs
    \item We evaluate the proposed methods over four standard IR benchmarks, demonstrating significant relative improvements vs. recent state of the art, of up to 18\% on nDCG@10 in pre-retrieval settings, and of up to 9\% nDCG@10 on post-retrieval (feedback) settings, demonstrating increased generalizability of our approach. Further analysis shows how performance is influenced by the number of feedback documents, the number of instructions and domain-specific instructions, and the generation of fluent reformulations.  
\end{itemize}
Next, we summarize the prior work to place our contributions in context.
\begin{figure*}[ht]
  \includegraphics[width=\textwidth]{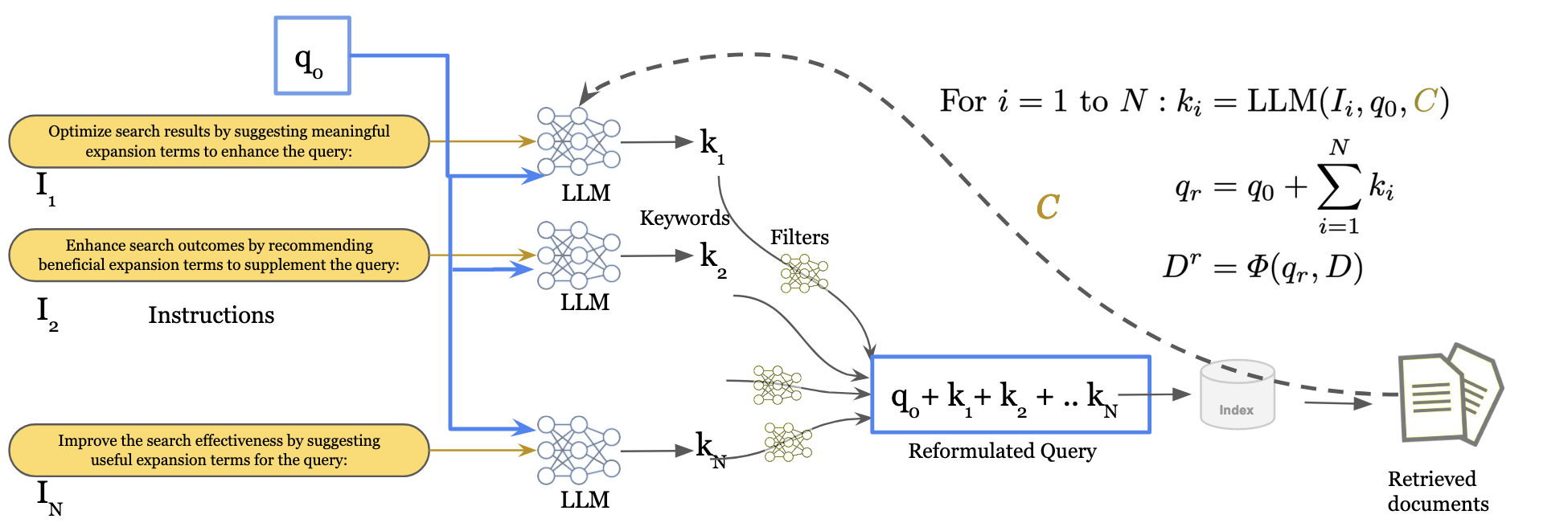}
  \caption{The complete flow and algorithm of \NameQR{} and \NamePRF{} (dotted).}
  \label{fig:unifiedquery}
\end{figure*}
\begin{figure*}[ht]
  \includegraphics[width=\textwidth]{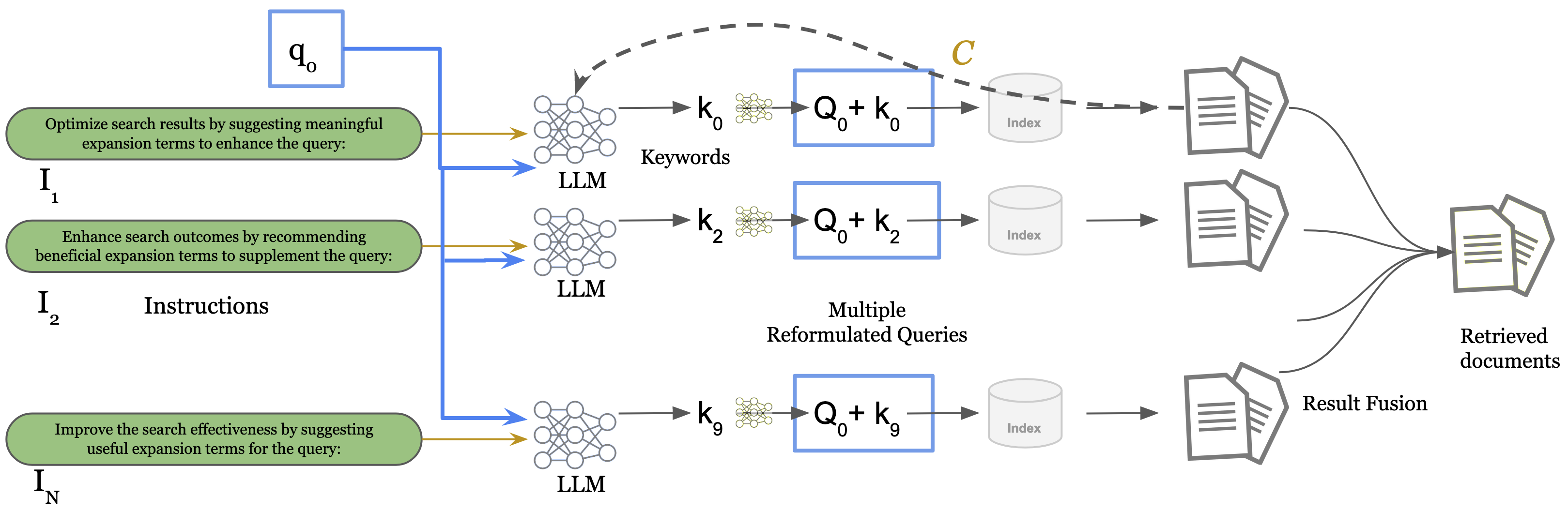}
  \caption{The complete flow of \NameDF{}. \NameDFPRF{} is shown with the dotted line.}
  \label{fig:docfusion}
\end{figure*}
\section{Related Work}
Query reformulation is effective in many settings~\cite{qrsurvey}. It can be used pre-retrieval, or post-retrieval, via incorporating evidence from feedback, obtained either from a user (relevance feedback) or from top-ranked results (pseudo-relevance feedback) in both sparse~\cite{prfsurvey} and dense retrieval settings~\cite{colbertprf,anceprf}. 

Recently, zero-shot approaches to query reformulation have received considerable attention.~\citet{zeroshot} design a query reformulator by zero-shot prompting an instruction tuned model, FlanT5~\cite{flant5} to generate keywords for query expansion and Pseudo-Relevance Feedback (PRF) incorporation.~\citet{promptingqueryexpa} demonstrate that LLMs can be more powerful than traditional methods for query expansion.~\citet{convsearch} propose a framework to reformulate conversational search queries using LLMs.~\cite{hyde}'s framework performs retrieval through fake documents generated by prompting LLMs with user queries.~\citet{dhole2023interactive,dhole2024queryexplorer} demonstrate an interactive zero-shot query generation and reformulation interface.~\citet{grf} vary the types of keywords to be generated by prompting for entities, news articles, essays, etc.~\citet{dhole2024genqrensemble} introduce~\NameQR{} by leveraging an ensemble of multiple prompts to generate a combined reformulation.

However, using a single query reformulation can often degrade performance compared to the original query. To address this drawback, prior efforts have incorporated ensemble strategies via keywords from numerous sources or fusing documents from different queries.~\citet{qu2} combine features extracted from various translation models to generate better query rewrites.~\citet{qu6} perform QR by utilizing multiple external biomedical resources.~\citet{qu7} present a data fusion framework suggesting that diverse query formulations represent distinct evidence sources for inferring document relevance. Later,~\citet{qu1} generated diverse queries by introducing a diversity-driven reinforcement learning algorithm. For other tasks, recent works demonstrated the benefits of ensemble strategies for prompting LLMs, including self-consistency~\cite{selfcons} for arithmetic and common sense tasks, Chain of Verification~\cite{cove} for improving factuality, and Diverse~\cite{diverse} for question answering. 

However, zero-shot based ensemble methods for LLMs have hardly been explored for the Query Reformulation task, as we propose in this paper.~\citet{grf} manually create different instructions by invoking different kinds of textual units i.e. news, documents, keywords, essays, etc., and find that combining keywords from above all improves retrieval performance in a traditional setting.~\citet{Li2023GenerateFA} describe an ensemble approach by splitting individual keywords generated from a single LLM pass to generate multiple reformulations and show their benefits for the cross-encoder based re-rankers.

\section{Proposed Approaches}
In this section, we describe two variations of our proposed approach, pre- and post-retrieval settings. In the pre-retrieval setting, $q_r = R.q_0$, where a Query Reformulation $R$ transforms a user's expressed query $q_0$ into a novel reformulated version $q_r$ to improve retrieval effectiveness for a given search task (e.g., passage or document retrieval). We also consider the post-retrieval setting, wherein the reformulator can incorporate additional context like document or passage-level feedback. 

\subsection{Pre-retrieval}
We introduce \textbf{\NameQR{}} and \textbf{\NameDF{}} -- two ensemble prompting-based query reformulators that use $N$ diverse paraphrases of a QR instruction to enhance retrieval. Specifically, we first use an LLM to paraphrase the instruction $I_1$ to create $N$ instructions with different surface forms viz. $I_1$ to $I_{N}$. This is done offline once. Each instruction is then prompted along with the user's query $q_0$ to generate instruction-specific keywords. These keywords are then optionally passed to another LLM to filter out irrelevant keywords.
\begin{enumerate}
    \item In \NameQR{}, all the keywords are then appended to the original query, resulting in a reformulated bag-of-words query, which is then executed against a document index $D$ to retrieve relevant documents $D^{r}$. The complete process and algorithm are shown in Figure~\ref{fig:unifiedquery}.
    \item In \NameDF{}, each of the multiple keywords generated from the individual instructions are appended one by one to create $N$ reformulations. These are then executed against a document index $D$ to produce $N$ sets of relevant documents $D_{i}$. The sets are then fused to create a single ranked list of documents $D^{r}$ (\textit{e.g.} score fusion between BM25 scores or reciprocal rank fusion). The complete process is shown in Figure~\ref{fig:docfusion}.
\end{enumerate}

\subsection{Post-retrieval}
To assess how well our method can incorporate additional context like document feedback, we introduce \textbf{\NamePRF{}} and \textbf{\NameDFPRF{}}. Here, we prepend the $N$ instructions described earlier with a fixed context capturing string ``\bluetexttt{Based on the given context information \{C\},}'' used\footnote{We found prepending this string performs better than appending it at the end during prompting} in ~\citet{zeroshot} to create their PRF counterparts -- where \bluetexttt{C} is a space (` ') delimited concatenation of feedback documents \bluetexttt{C} $= d_1 + \ldots + d_m$, obtained either as pseudo-relevance feedback from first-stage retrieval, or human-feedback. 
\section{Experiments}\label{experiments}
We now describe the experiments and analysis performed for different retrieval settings. 
\subsection{Prompts}
To instruct the LLM to generate query reformulations, we start with an instruction empirically chosen by~\citet{zeroshot} -- as our base QR instruction $I_1$. We use this instruction to generate $N$ paraphrases ($N=10$). To this aim, we invoke the~\href{https://chat.openai.com/}{ChatGPT} API~\cite{openai2023gpt4} with the paraphrase generating prompt, namely, $I_p$ = \bluetexttt{``Generate 10 paraphrases for the following instruction:''}-- and the base QR instruction $I_1$ to obtain $I_2$ to $I_{10}$. These paraphrases, shown in Figure~\ref{fig:paraphrases} serve as our instruction set for all subsequent experiments.

We further resort to a domain-specific instruction for each of the datasets described in the upcoming section which is paraphrased similarly.
\begin{figure}[h]
\centering
\resizebox{\columnwidth}{!}{%
\begin{tabular}{ll}
\# & ~\textbf{Instruction}\\
\hline
1 &  Improve the search effectiveness by suggesting expansion terms for the query\\
2 &  Recommend expansion terms for the query to improve search results\\
3 &  Improve the search effectiveness by suggesting useful expansion terms for the query\\
4 &  Maximize search utility by suggesting relevant expansion phrases for the query\\
5 &  Enhance search efficiency by proposing valuable terms to expand the query\\
6 &  Elevate search performance by recommending relevant expansion phrases for the query\\
7 &  Boost the search accuracy by providing helpful expansion terms to enrich the query\\
8 &  Increase the search efficacy by offering beneficial expansion keywords for the query\\
9 &  Optimize search results by suggesting meaningful expansion terms to enhance the query\\
10 &  Enhance search outcomes by recommending beneficial expansion terms to supplement the query
\end{tabular}
}
\caption{The $N$ reformulation instructions used for \NameQR{} and \NameDF{}}
\label{fig:paraphrases}
\end{figure}
\subsection{Generation Models}
For generating the query reformulations, we employ two models,~\bluetexttt{flan-t5-xxl} and ~\bluetexttt{Llama-2-7b-chat-hf}. FlanT5~\cite{flant5} is a set of models are created by fine-tuning the text-to-text transformer model,~\bluetexttt{T5}~\cite{t5} on instruction data of a variety of NL tasks. We use the checkpoint\footnote{\href{https://huggingface.co/google/flan-t5-xxl}{\texttt{huggingface.co/google/flan-t5-xxl}}} provided through HuggingFace’s Transformers library(HF)~\cite{huggingface}. Nucleus sampling is performed with a cutoff probability of 0.92 keeping top 200 tokens (top\_k) and a repetition penalty of 1.2.

We also investigate the use of ~\bluetexttt{Llama-2-7b-chat-hf}~\cite{touvron2023llama}, an auto-regressive language model, which is RLHF fine-tuned and optimized for dialog use cases. We chose the LLama2 series of models as they have shown state-of-the-art performance across multiple benchmarks. We use the HF checkpoint\footnote{\href{https://huggingface.co/meta-llama/Llama-2-7b-chat-hf}{\texttt{huggingface.co/meta-llama/Llama-2-7b-chat-hf}}} keeping the same generation settings as above with a repetition penalty of 2.1. We use the prompt template shown in Figure~\ref{fig:llamaprompt} where the \texttt{instruction} variable is the concatenation of the actual instruction and the query provided at run-time. We appended ``And do not explain yourself.'' to minimize the conversational jargon that the model could generate.

\begin{figure}[h]
\centering
\resizebox{\columnwidth}{!}{%
\fbox{
\begin{tabular}{l}
You are a helpful assistant who directly provides \\
     comma separated keywords or expansion terms. \\
     Provide as many expansion terms or keywords as possible \\
     related to the query. And do not explain yourself.\\
     \texttt{instruction}: \texttt{query}
\end{tabular}
}
}
\caption{The prompt used for all the Llama-2 Query reformulators.}
\label{fig:llamaprompt}
\end{figure}

\begin{table*}[!ht]
    \centering
    \resizebox{\textwidth}{!}{ %
    \begin{tabular}{llll|lll|ll|ll}
     ~ &~\textbf{Reformulation} &
      \multicolumn{2}{c}{ \textbf{TREC Passage 19} } &
      \multicolumn{3}{c}{ \textbf{TREC Robust 04} } &
      \multicolumn{2}{c}{ \textbf{Webis Touche} } &
      \multicolumn{2}{c}{ \textbf{DBpedia Entity} } \\ \hline
        Model & Name & nDCG@10 & RR(rel=2) & P@10 & nDCG@10 & RR(rel=2) & nDCG@10 & RR(rel=2) & nDCG@10 & RR(rel=2) \\ 
        ~ & BM25 & .480 & .642 & .426 & .434 & .154 & .260 & .454 & .321 & .297 \\
        \texttt{GPT3-curie} & Query2Doc~\cite{wang-etal-2023-query2doc} & .551 & - & - & - & - & - & - & - & - \\ \hline
       \multirow{9}{*}{\texttt{flan-t5-xxl}} & GenQR & .477 & .593 & .473 & .483 & .151 & .315 & .511 & .342 & .345 \\ 
      ~ & GenQR$_{\beta=.05}$ & .511 & .621 & .469 & .477 & .150 & .276 & .476 & .353 & .339 \\ 
      ~ & GenQR+FL & .489 & .653 & .439 & .446 & .151 & .262 & .459 & .326 & .305\\ 
      ~ & GenQR~\cite{zeroshot} & .556 & .707 & - & .461 & - & - & - & - & -\\ \cmidrule{2-11}
        ~ & \NameQR{}~\cite{dhole2024genqrensemble} & .564$^\dagger$ & .706 & .500$^\dagger$ & \textbf{.513}$^\dagger$ & .159 & \textbf{.317} & \textbf{.555} & .374$^\dagger$ & .376$^\dagger$ \\ 
                ~ & \NameQR{}$_{\beta=.05}$ & \textbf{.575}$^\dagger$ & .714 & \textbf{.502}$^\dagger$ & .512$^\dagger$ & .159 & .292 & .489 & \textbf{.377}$^\dagger$ & \textbf{.380}$^\dagger$ \\
                ~ & \NameQR{}+FL & .537$^\dagger$ & .694 & .482$^\dagger$ & .492$^\dagger$ & .151 & .272 & .467 & .361$^\dagger$ & .341$^\dagger$\\
        ~ & \NameDF{} & .565$^\dagger$ & \textbf{.717}$^\dagger$ & \textbf{.502}$^\dagger$ & \textbf{.513}$^\dagger$ & \textbf{.160} & .302 & .529 & .373 & .379 \\ 
        ~ & \NameDF{}+FL & .558 & .698$^\dagger$ & .493 & .503 & .155 & .278 & .489 & .370$^\dagger$ &.360  \\ \hline
        \multirow{7}{*}{\texttt{llama-2-7b}} & GenQR & .496 & .678 & .451 & .461 & .156 & .268 & .462 & .328 & .304 \\ 
         & GenQR+FL & .493 & .666 & .449 & .459 & .156 & .267 & .464 & .329 & .304 \\
         & KEQE~\cite{lei2024corpus} & .571 & - & - & - & - & - & - & - & - \\ \cmidrule{2-11}
        ~ & \NameQR{} & \textbf{.580}{$^\dagger$} & .703$^\dagger$ & .504$^\dagger$ & .513$^\dagger$ & \textbf{.173} & \textbf{.319}{$^\dagger$} & \textbf{.579}{$^\dagger$} & .370{$^\dagger$} & \textbf{.372}{$^\dagger$} \\ 
        & \NameQR{}+FL & .575$^\dagger$ & \textbf{.744} & \textbf{.510}$^\dagger$ & \textbf{.519}$^\dagger$ & .166 & .314$^\dagger$ & .543 & .\textbf{.372}$^\dagger$ & \textbf{.372}$^\dagger$ \\ 
        ~ & \NameDF{} & .503 & .679 & .453 & .461 & .158 & .268 & .465 & .328 & .304 \\
        ~ & \NameDF{}+FL & .497 & .673 & .448 & .457 & .157 & .268 & .463 & .326 & .304 \\
    \end{tabular}}
     \caption{Performance of \NameQR{} and \NameDF{} on the four benchmarks with two underlying models. $^\dagger$ denotes significant improvements (paired t-test with Holm-Bonferroni correction, $p<0.05$) over GenQR.}
\label{ensembleperformance_bm25}
\end{table*}
\subsection{Retrieval Evaluation}
For evaluation, we use four popular benchmarks through IRDataset~\cite{irds}'s interface.

1)~\textbf{TP19}: TREC 19 Passage Ranking which uses the MSMarco dataset~\cite{msmarco,promptingqueryexpa} consisting of search engine queries. 2)~\textbf{TR04}: TREC Robust 2004 Track, a task intended for testing poorly performing topics. In our experiments, we use~\texttt{title} as our choice of query. And two tasks from the BEIR~\cite{beir} benchmark 3)~\textbf{WT}: Webis Touche~\cite{touche2020} for argument retrieval and 4)~\textbf{DE}: DBPedia Entity Retrieval~\cite{dbpedia}, a test collection for entity search.
\subsection{Baselines} 

We compare our work against the following using the Pyterrier~\cite{pyterrier} framework. For all the post-retrieval experiments, we used 5 documents as feedback.

\subsubsection{Experiments with BM25 Retriever}
Here, we compare with the following approaches. 
\begin{enumerate}
    \item BM25: Here, we retrieve using raw queries without any reformulation
    \item GenQR: We implement a single-instruction zero-shot QR~\cite{zeroshot} which is also a specific case of our approach when N=1 with~\texttt{flan-t5-xxl} and~\texttt{llama-2-7b}. 
    \item Query2Doc~\cite{wang-etal-2023-query2doc}'s reformulation generated via a~\texttt{GPT3-curie} model (6.7B), which is the closest in size to ~\texttt{flan-t5-xxl} (11.3B) and~\texttt{llama-2-7b} (7B).
    \item BM25+RM3~\cite{rm3}: BM25 retrieval with RM3 expanded queries (\#feedback terms=10)
    \item BM25+GenPRF~\cite{zeroshot}: BM25 retrieval with GenPRF expanded queries
    \item \NameQR{}: The ensemble query reformulation introduced by~\citet{dhole2024genqrensemble} which used a FlanT5 generator.
    \item \NamePRF{}: The corresponding PRF variant.
\end{enumerate}

\subsubsection{Experiments with Neural Reranking} Here, we re-evaluate the above settings in conjunction with a MonoT5 neural reranker with all other parameters constant. We use the MonoT5 base version~\texttt{castorini/monot5-base-msmarco} through the~\texttt{PyTerrier\_t5} plugin\footnote{\url{https://github.com/terrierteam/pyterrier_t5}}.

\begin{enumerate}
    \item BM25+MonoT5: BM25 retrieval using raw queries, re-ranked with MonoT5 model~\cite{monoT5}  
    \item \{\NameQR{}$\backslash$GenQR\}+MonoT5: BM25 retrieval with \NameQR{}$\backslash$ GenQR reformulations, re-ranked with MonoT5 model
    \item BM25+RM3+MonoT5: BM25 retrieval with RM3 expanded queries, re-ranked with MonoT5 model
    \item BM25+\{\NamePRF{}$\backslash$GenPRF\}+MonoT5: BM25 retrieval with \NamePRF{}$\backslash$ GenPRF expanded queries, re-ranked with MonoT5 model
\end{enumerate}

\begin{table*}[!ht]
    \centering
    \resizebox{\textwidth}{!}{ %
    \begin{tabular}{clll|lll|ll|ll}
     ~ &~\textbf{Reformulation} &
      \multicolumn{2}{c}{ \textbf{TREC Passage 19} } &
      \multicolumn{3}{c}{ \textbf{TREC Robust 04} } &
      \multicolumn{2}{c}{ \textbf{Webis Touche} } &
      \multicolumn{2}{c}{ \textbf{DBpedia Entity} } \\ \hline
        Model & Name & nDCG@10 & RR(rel=2) & P@10 & nDCG@10 & RR(rel=2) & nDCG@10 & RR(rel=2) & nDCG@10 & RR(rel=2) \\ 
        ~ & BM25+MonoT5 & .718 & .881 & .492 & .513 & .173 & .299 & .525 & .414 & .444 \\ \hline
\texttt{GPT3-curie} & Query2Doc~\cite{wang-etal-2023-query2doc} & .687& - & - & - & - & - & - & - & -\\
 - & HyDE~\cite{hyde} & .613& - & - & - & - & - & - & - & -\\ 
 - & Doc2Query~\cite{doc2query} & .627 & - & - & - & - & - & - & - & -\\
 - & Doc2Query--~\cite{gospodinov2023doc2query} & .670 & - & - & - & - & - & - & - & -\\
 ~ & T5QR~\cite{zeroshot} & .696 & .831 & - & .474 & - & - & - & - & - \\
 \hline
      \multirow{6}{*}{\texttt{flan-t5-xxl}} & GenQR$^{M}$ & .707 & .847 & .490 & .510 & .170 & .292 & .530 & .415 & .446 \\
      ~ & GenQR$^{M}${}\textnormal{+FL} & .720 & .881 & .491 & .511 & .170 & .299 & .530 & .415 & .438 \\ 
      ~ & GenQR$^{M}${}\textnormal{~\cite{zeroshot}} & .727 & .908 & ~ & .473 & - & - & - & - & - \\ \cmidrule{2-11}
        ~ & \NameQR{}$^{M}${}\textnormal{~\cite{dhole2024genqrensemble}} & .722 & .862 & \textbf{.484} & \textbf{.506} & \textbf{.170} & .298 & \textbf{.548} & \textbf{.420} & .450 \\
        ~ & \NameQR{}$^{M}$\textnormal{+FL} & .725 & .867 & .488 & .509 & .170 & .297 & .528 & .418 & .438 \\
        ~ & \NameDF{}$^{M}$ & \textbf{.723} & \textbf{.875} & .404 & .422 & .157 & \textbf{.301} & .529 & .403 & \textbf{.433} \\ 
        ~ & \NameDF{}$^{M}$\textnormal{+FL} & .729 & .881 & .438 & .455 & .163 & .296 & .528 & .411 & .439 \\ \hline
        \multirow{6}{*}{\texttt{llama-2-7b}} & GenQR$^{M}$ & .729 & \textbf{.881} & \textbf{.490} & \textbf{.510} & \textbf{.170} & \textbf{.300} & .530 & .414 & .442  \\ 
        ~ & GenQR$^{M}$\textnormal{+FL} & .719 & .858 &.488 & .510 & .170 & \textbf{.300} &.528 & \textbf{.419} & \textbf{.448}  \\ \cmidrule{2-11}
        ~ & \NameQR{}$^{M}$ & \textbf{.730} & .869 & .488 & .510 & \textbf{.170} & \textbf{.300} & \textbf{.532} & \textbf{.415} & \textbf{.444} \\
        ~ & \NameQR{}$^{M}$\textnormal{+FL} & .729 & .869 & \textbf{.490} & \textbf{.511} & .169 & .299 & .527 & .415 & .441 \\
        ~ & \NameDF{}$^{M}$ &.728 & \textbf{.881} & .470 & .487 & .168 & \textbf{.300} & .528 & .412 & .440 \\ 
        ~ & \NameDF{}$^{M}$\textnormal{+FL} & .721 & .881 & .470 & .488 & .166 & \textbf{.300} & .528 & .413 & .441 \\ 
    \end{tabular}}
     \caption{\blt{Performance of \NameQR{} and \NameDF{} under the neural setting and compared with other neural approaches. The superscript M denotes reranking through a MonoT5 reranker.}}
\label{ensembleperformance_neural}
\end{table*}

\section{Results and Analysis}
We now report the results of query reformulation for all the settings and present further analysis.
\subsection{Pre-Retrieval Performance}
\begin{figure*}
    \centering
    \fbox{
    \includegraphics[width=1\textwidth]{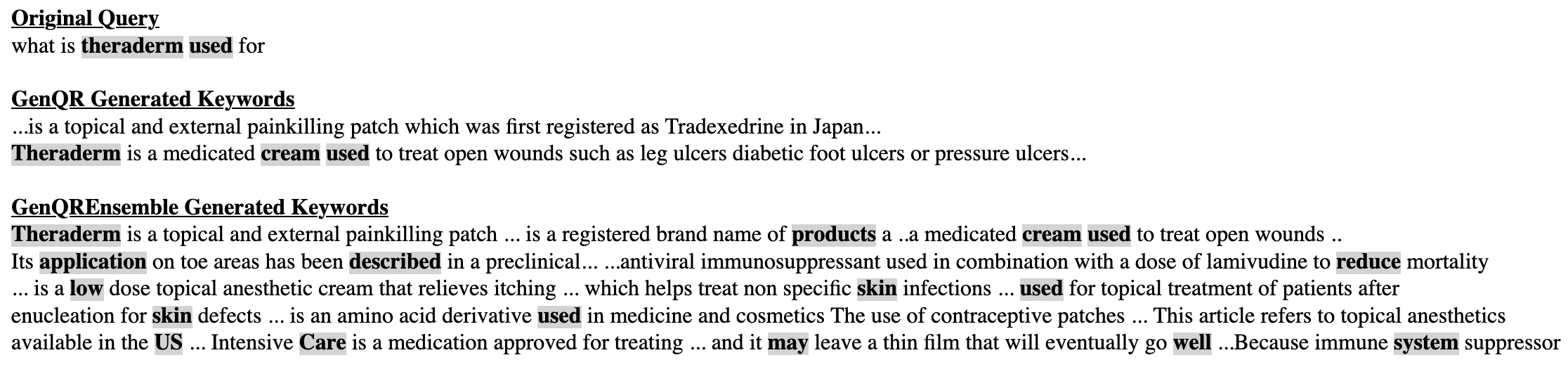}}
    \caption{Query Reformulations generated from the Single Instruction Setting and the Ensemble Setting. The grey highlights depict the terms present in the highest relevant (gold) documents.}
    \label{fig:sample_example_grey}
\end{figure*}

\begin{figure*}[htbp]
\centering
\begin{minipage}{0.5\textwidth}
  \centering
  \includegraphics[width=1\linewidth]{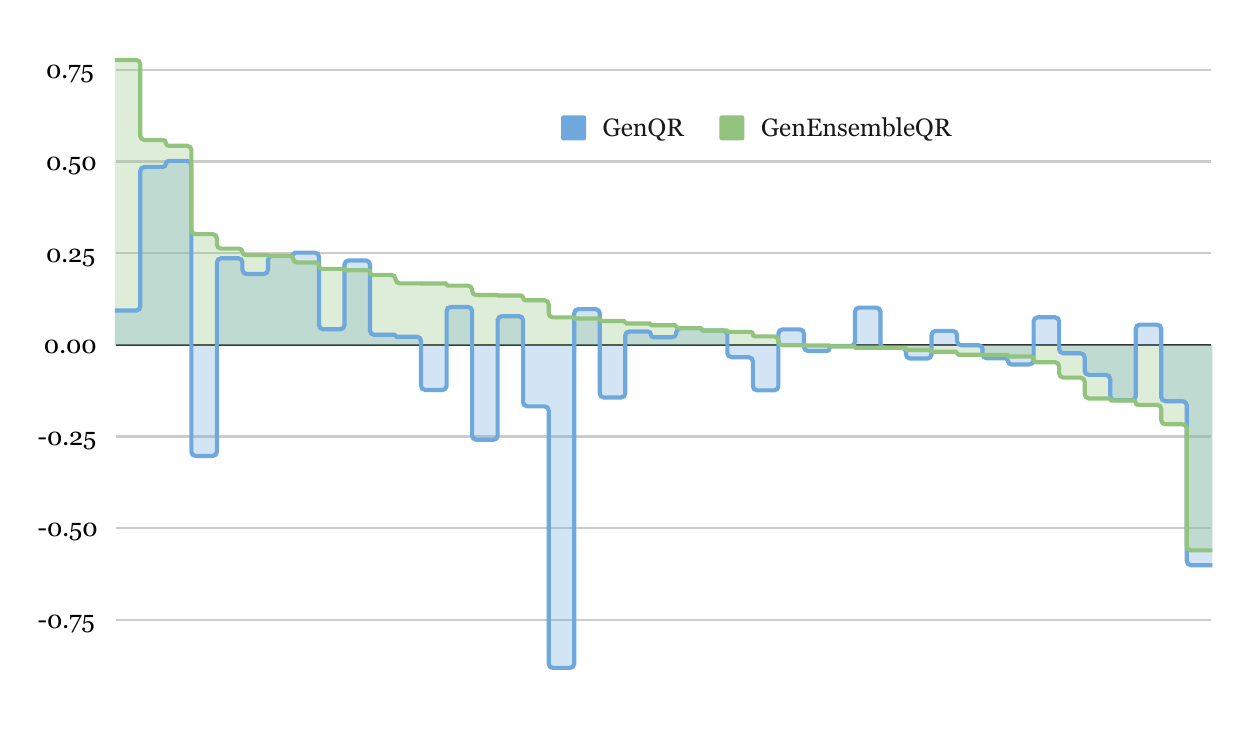}
\end{minipage}%
\begin{minipage}{0.5\textwidth}
  \centering
    \includegraphics[width=1\linewidth]{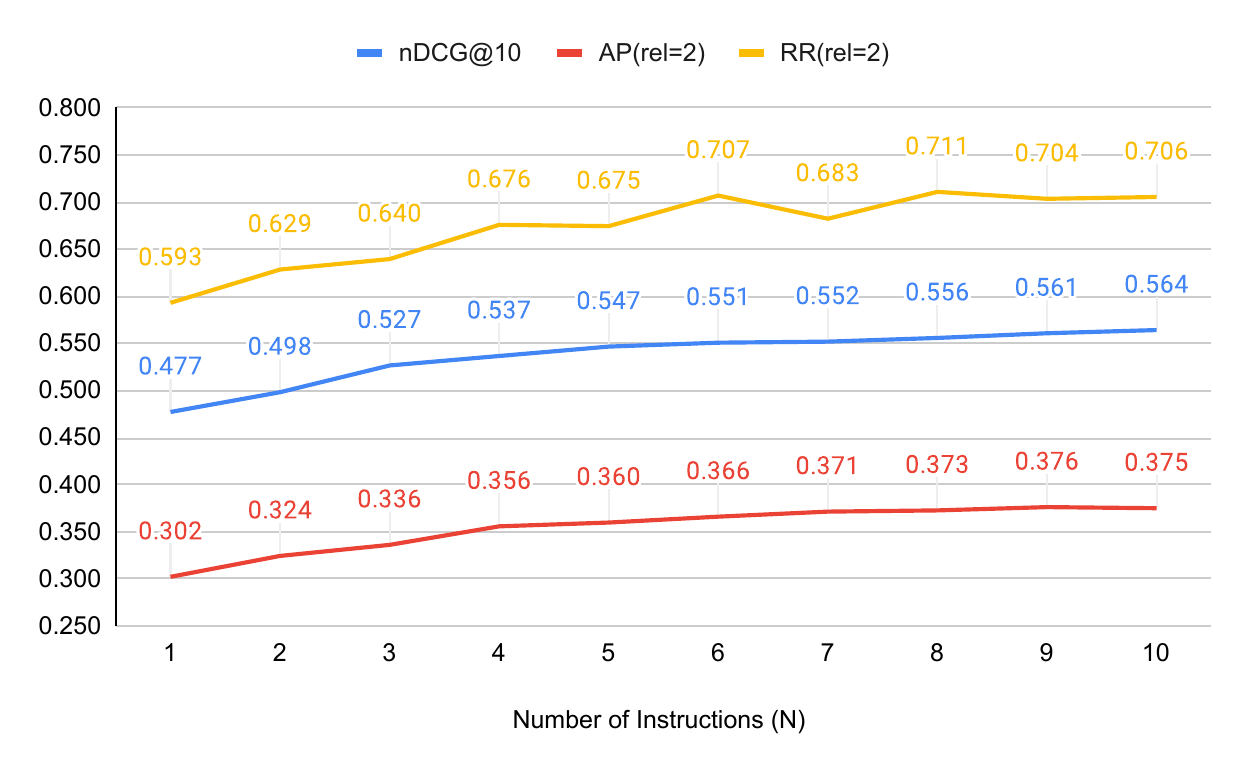}
\end{minipage}
\caption{1) Relative nDCG@10 scores of \NameQR{} as compared to GenQR on TP19 benchmark 2) On TP19, \NameQR{} improves as the number of instructions are increased.}
\label{fig:3-analysis}
\end{figure*}

We first compare the retrieval performances of raw queries and the reformulations from GenQR, with \NameQR{} and \NameDF{} in Table~\ref{ensembleperformance_bm25}. Both \NameQR{} and \NameDF{} outperform the raw queries as well as generate better reformulations than GenQR's queries across all four benchmarks over a BM25 retriever. ~\NameQR{} improves performance over the single instruction setting through keywords generated from both the underlying generators, FlanT5 and Llama-7-b with Llama-7-b showing better gains. The performance also gradually improves with more number of instructions as seen in Figure~\ref{fig:3-analysis}.2. This indicates the usefulness of paraphrasing initial instructions and exploiting the model's sensitivity. On TP19, nDCG@10 and MAP improve significantly with relative improvements of 18\% and 24\% respectively. This is further validated through the querywise analysis shown in Figure~\ref{fig:3-analysis}.1 -- Relative to BM25, nDCG@10 scores of \NameQR{} are overall better than GenQR.\NameQR{} seems more robust too as it avoids drastic degradation in at least 6 queries on which GenQR fails. Example reformulations are shown in Figure~\ref{fig:sample_example_grey}, where crucial keywords are produced through \NameQR{}'s paraphrasing.

We further look at \NameQR{} under the neural reranker setting shown in Table~\ref{ensembleperformance_neural}. In three of the four settings, viz., TP19, WT, and DE, \NameQR{} is preferable to its zero-shot variant, GenQR. Evidently, the gains of both the zero-shot approaches in the traditional setting are stronger vis-à-vis the neural setting. We hypothesize this could be due to \NameQR{} and GenQR both expanding the query by incorporating semantically similar but lexically different keywords. Comparatively, neural models are adept at capturing notions of semantic similarity and might benefit less from QR. This also is in line with~\citet{fails}'s recent analysis on the non-ensemble variant.

\subsection{Post-Retrieval Performance}

\begin{table*}[htbp]
    \centering
\resizebox{\textwidth}{!}{%
    \begin{tabular}{lllll|llll}
      &
      \multicolumn{4}{c}{ \textbf{With BM25 Retriever} } &
      \multicolumn{4}{c}{ \textbf{With Neural Reranking} } \\ \hline
        Setting & nDCG@10 & nDCG@20 & MAP & MRR & nDCG@10 & nDCG@20 & MAP & MRR \\
        BM25 & .480 & .473 & .286 & .642 & .718 & .696 & .477 & .881 \\ 
        RM3 & .504 & .496 & .311 & .595 & .716 & .699 & .480 & .858 \\ \hline
        GenPRF & .576 & .553 & .363 & .715 & .722 & .703 & .486 & .874 \\ 
        GenPRF~\cite{zeroshot} & \textbf{.628} & - & \textbf{.404} & .809 & - & - & - & - \\
        \NamePRF{} & \textbf{.585}\relup{2} & \textbf{.560}\relup{1} & \textbf{.373}\relup{3} & \textbf{.753}\relup{5} & \textbf{.729}\relup{1} & \textbf{.706}\relup{1} & \textbf{.501}\relup{3} & \textbf{.894}\relup{2} \\ 
        \NameDFPRF{} & .566 & .548 & .368 & .725 & .718 & .707 & .488 & .882 \\
        \hline 
        GenPRF (Oracle) & .753 & .728 & .501 & .936 & .742 & .734 & .545 & .881 \\ 
        \NamePRF{} (Oracle) & \textbf{.820}$^\alpha$\relup{9} & \textbf{.773}\relup{6} & \textbf{.545}\relup{9} & \textbf{.977}\relup{4} & \textbf{.756}\relup{2} & \textbf{.751}\relup{2} & \textbf{.545} & \textbf{.897}\relup{2} \\
        \NameDFPRF{} (Oracle) & .708 & .672 & .465 & .938 & .748 & .731 & .532 & .887 \\
    \end{tabular}}
    \caption{\blt{Comparison of PRF performance on the TP19 benchmark using queries generated from \texttt{flan-t5-xxl}}}
    \label{prf_results}
\end{table*}
We now investigate if \NamePRF{} and \NameDFPRF{} can effectively incorporate PRF in Table~\ref{prf_results}. While \NameDFPRF{} improves recall, we find that \NamePRF{} improves retrieval performance across all metrics as compared to other PRF approaches and is able to incorporate feedback from a BM25 retriever better than RM3 as well as its zero-shot counterpart. To assess if \NamePRF{} and GenPRF can at all benefit from incorporating relevant documents, we perform oracle testing by providing the highest relevant gold documents as context. We find that \NamePRF{} is able to improve over \NameQR{} (without feedback) showing that it is able to capture context effectively as well as benefit from it. It can incorporate relevant feedback better than its single-instruction counterpart GenPRF. We notice improvements even under the neural setting as \NamePRF{} outperforms RM3 and GenPRF. Besides, the oracle improvements are higher with only a BM25 retriever as compared to when a neural reranker is introduced.

\begin{figure*}[htbp]
\centering
\begin{minipage}{0.48\textwidth}
  \centering
    \includegraphics[width=\columnwidth]{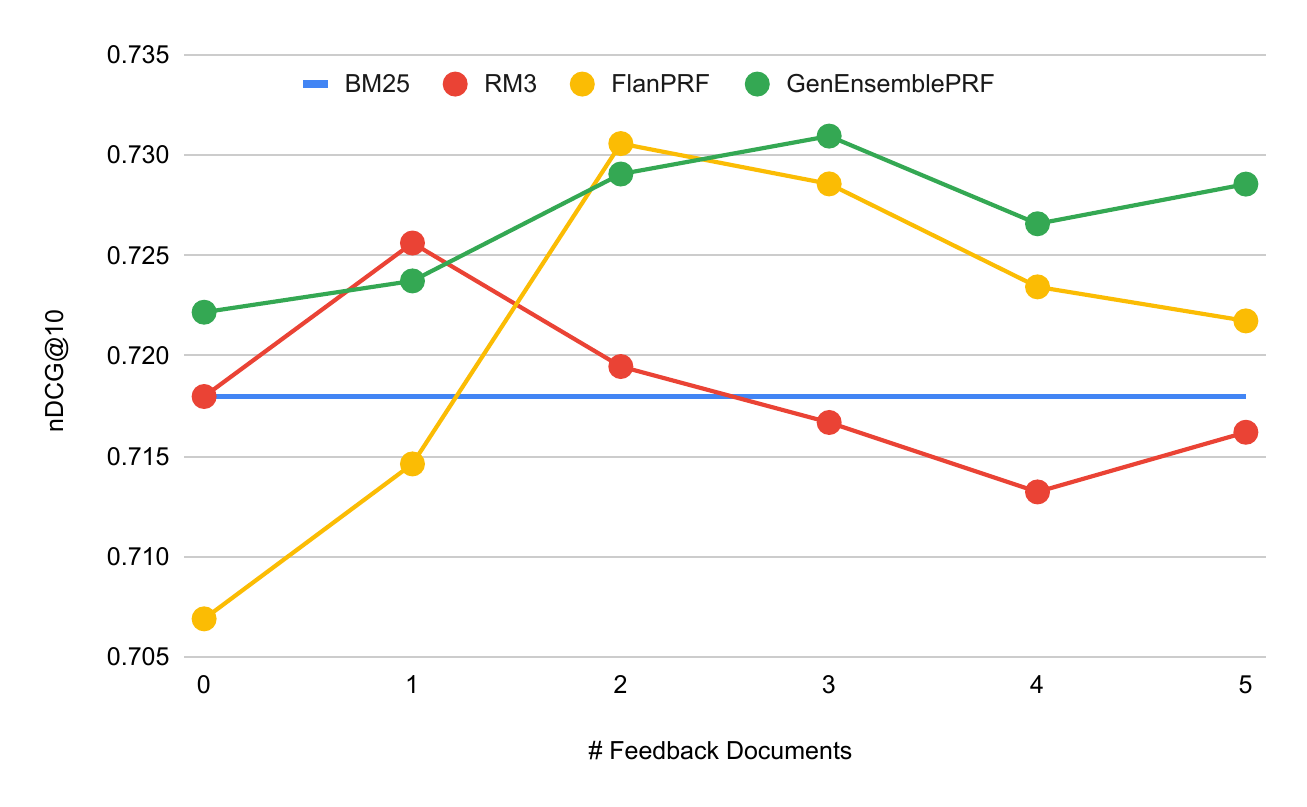}
\end{minipage}
\begin{minipage}{0.48\textwidth}
  \centering
    \includegraphics[width=\columnwidth]{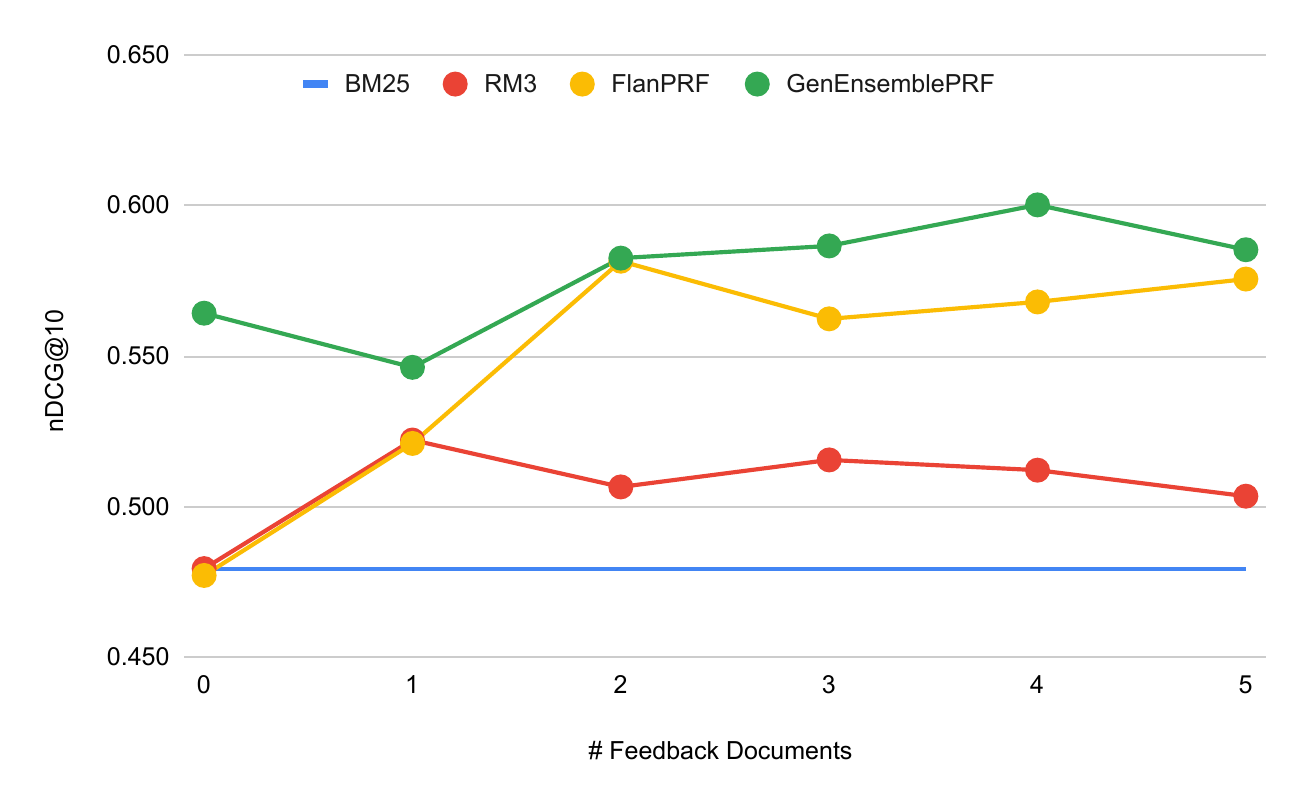}
\end{minipage}
\caption{Effect of PRF with increasing feedback on the sparse (BM25) and the neural (MonoT5) setting on TP19.}
\label{fig:increasing_feedback}
\end{figure*}

\section{Analysis}
Our complete ensemble pipeline has multiple directions in which performance can be improved. In this section, we attempt to vary the number of feedback documents, relative influence of the keywords and incorporate domain specific instructions to seek additional gains. 
\begin{table*}[ht]
    \centering
    \resizebox{\textwidth}{!}{%
    \begin{tabular}{lll|ll|ll|ll}
     &
      \multicolumn{2}{c}{ \textbf{TREC Passage 19} } &
      \multicolumn{2}{c}{ \textbf{TREC Robust 04} } &
      \multicolumn{2}{c}{ \textbf{Webis Touche} } &
      \multicolumn{2}{c}{ \textbf{DBpedia Entity} } \\ \hline
        Setting & nDCG@10 & RR(rel=2) & nDCG@10 & RR(rel=2) & nDCG@10 & RR(rel=2) & nDCG@10 & RR(rel=2) \\ 
        GenQR & .496 & .678 & .461 & .156 & .268 & .462 & .328 & .304 \\ 
        +DSI & \textbf{.504} & \textbf{.679} & \textbf{.467} & \textbf{.158} & \textbf{.271} & \textbf{.471} & \textbf{.331} & \textbf{.309} \\ \hdashline
        GenEnsembleQR & .580 & .703 & .513 & .173 & .319 & .579 & .370 & .372 \\ 
        +DSI & .573 & .693 & \textbf{.522} & .169 & \textbf{.330} & .576 & \textbf{.377} & \textbf{.378} \\ \hline
        GenQR+MonoT5 & .729 & .881 & .510 & .170 & .300 & .530 & .414 & .442 \\ 
        +DSI & .723 & .881 & .509 & \textbf{.171} & .297 & .530 & \textbf{.417} & .437 \\ \hdashline
        GenEnsembleQR+MonoT5 & .730 & .869 & .510 & .170 & .300 & .532 & .415 & .444 \\ 
        +DSI & \textbf{.732} & .869 & \textbf{.512} & \textbf{.171} & .299 & .530 & \textbf{.416} & \textbf{.448} \\ 
    \end{tabular}}
    \caption{Impact of using domain-specific instruction (DSI) as the initial instruction. Values in bold signify where the domain-specific instruction improves performance vis-à-vis the general instruction counterpart.}
    \label{llama2DS}
\end{table*}
\begin{figure}[ht]
  \includegraphics[width=\columnwidth]{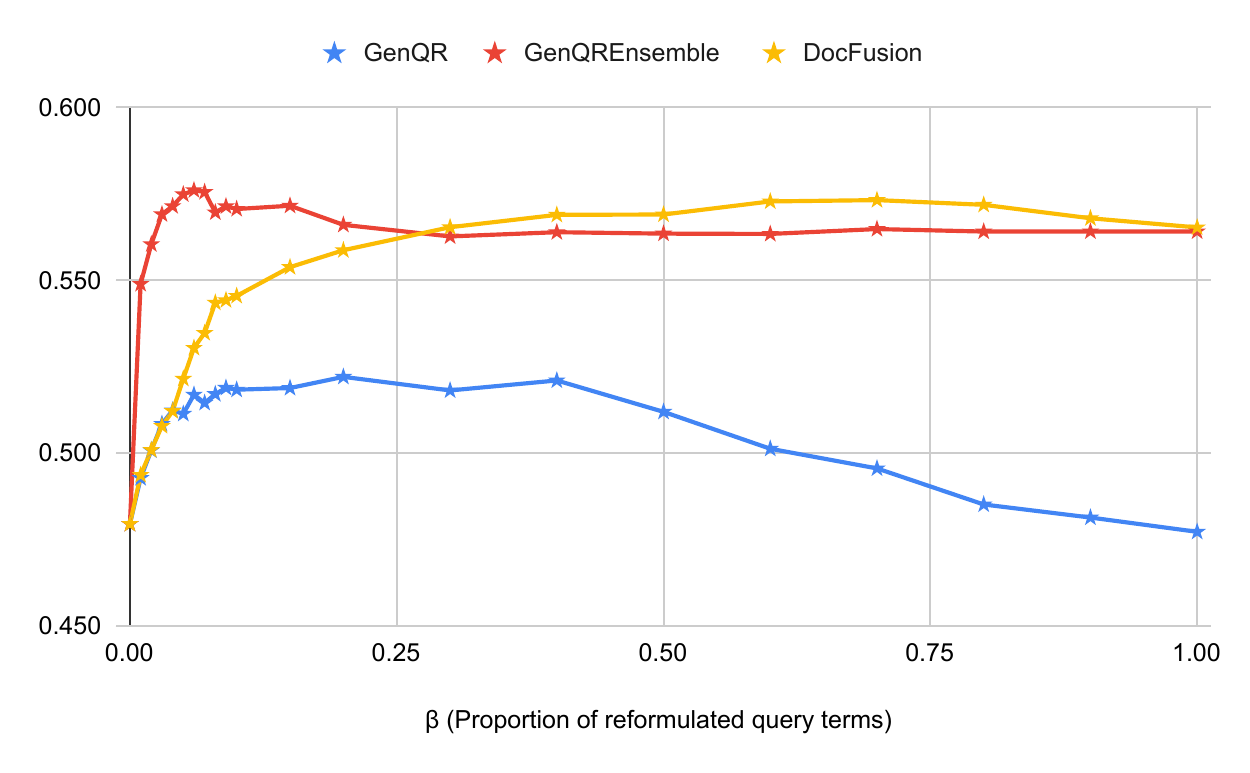}
  \caption{Effect on TP19 nDCG@10 on varying the relative influence of the reformulated query as compared to the original query.}
  \label{fig:reformweights}
\end{figure}
\subsubsection{Increasing Feedback Documents} We further evaluate the effect of varying the number of feedback documents from 0 to 5. We notice that resorting to an ensemble approach is highly beneficial as seen in Figure~\ref{fig:3-analysis}.2. In the BM25 setting, the ensemble approach seems always preferable. Under the neural reranker setting too,\NamePRF{} almost always outperforms GenPRF.
\begin{figure*}
\centering
\label{fig:FLtered_query}
\resizebox{\textwidth}{!}{%
\fbox{
\begin{tabular}{l}
\textbf{Reformulation}: 12Expansion termsof Gold fish growing growthrate food size tanks health Golden  Goldie s Aquatic Fish Food Growth Size Inquiry 1 size of golden fish 2 golden size limits \\admission tickets in an underwater setting   3  compilation records from aquatic plants about growth statistics 4 information on food required 1  Aquarium care tip sheet growth rate chart   ATSolutions \\MidCoastUtePlymDump FASTBACCP 13 keyworssgrowth  development increase in length aquarium pets water body size environment food diet care exercise breeding habitat lifespan \\
\textbf{Post Filtering}: Goldfish growing, growth rate, food, size, health, Aquatic Fish Food Growth Size, size of golden fish, information on food required, Aquarium care tip sheet growth rate chart,\\ growth development, increase in length, aquarium pets, water body size, environment, food, diet, care, breeding, habitat, lifespan.
\end{tabular}
}}
\caption{Sample Reformulation from~\texttt{flan-t5-xxl} for the query ``do goldfish grow'' before and after filtering}
\end{figure*}
\subsubsection{Relative Influence of Reformulation}
We vary the relative influence of the reformulated query by upweighting its terms as compared to the original query. We use the constant~\texttt{$\beta \in [0,1]$} to denote the proportional of reformulated query terms. When~\texttt{$\beta=0$}, the query consists of original query terms, and when~\texttt{$\beta=1$}, the query consists of terms from the original query as well as the generated keywords. We plot the nDCG@10 scores on MS-Marco for GenQR and \NameQR{} in Figure~\ref{fig:reformweights}. We notice that the performance peaks for~\texttt{$\beta$} values of less than 0.2. We use \texttt{terrierql} query language and linearly interpolate between the original query and the reformulated query. 

Besides, the performance of the single instruction setting tends to decrease with the increasing influence of reformulated terms while in the ensemble setting, the performance remains constant. This entails that ensemble prompting can not only produce more useful keywords but also reemphasize the crucial ones through different instructions. 
\begin{figure}
\centering
\resizebox{\columnwidth}{!}{%
\fbox{
\begin{tabular}{l}
Given a focused collection of arguments and some socially\\
     important and controversial topic, the keywords should\\
     retrieve arguments that could help an individual forming\\
     an opinion on the topic, or arguments that support/challenge\\
     their existing stance. Improve the search effectiveness of argument\\ retrieval by suggesting related expansion terms for the query.\\
\end{tabular}}
}
\caption{QR prompt used with~\texttt{llama-2-7b}}
\label{fig:domainprompt}
\end{figure}
\subsection{Incorporating Domain Information}
\begin{figure*}
\centering
\begin{minipage}{0.5\textwidth}
  \centering
  \includegraphics[width=1\columnwidth]{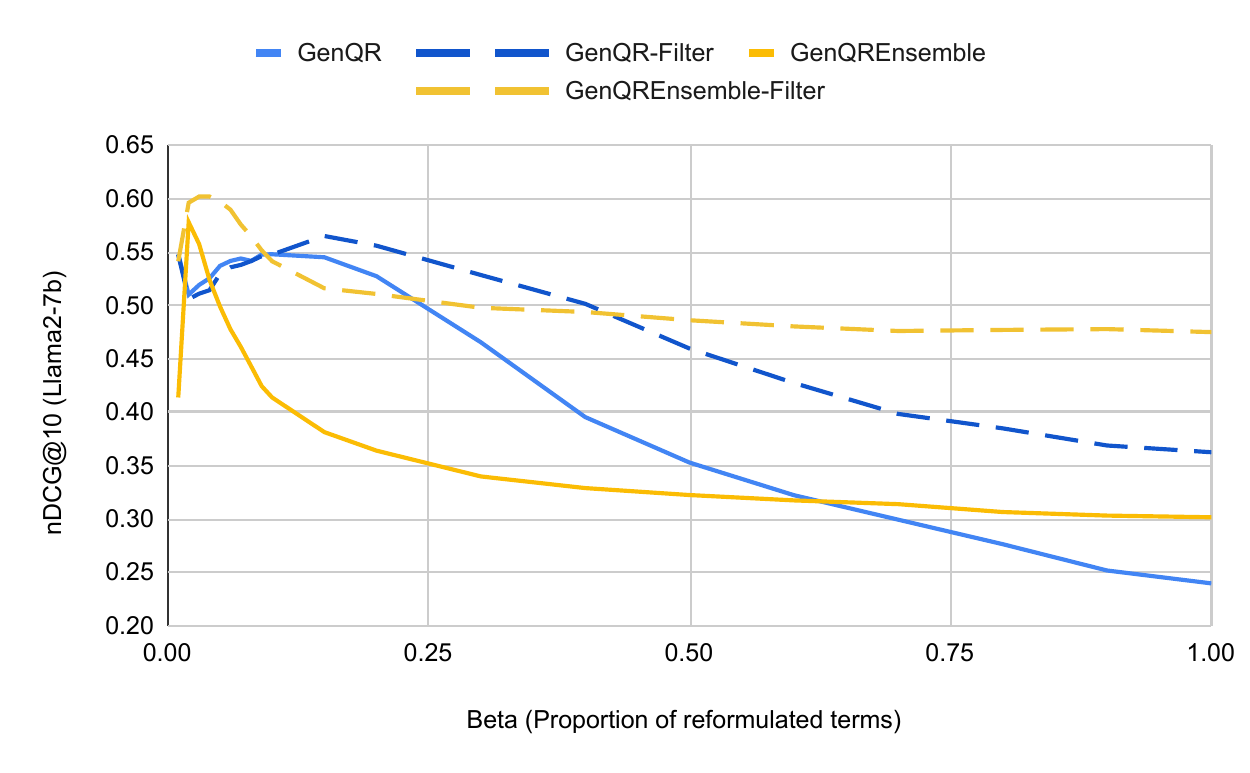}
\end{minipage}%
\begin{minipage}{0.5\textwidth}
  \centering
    \includegraphics[width=1\columnwidth]{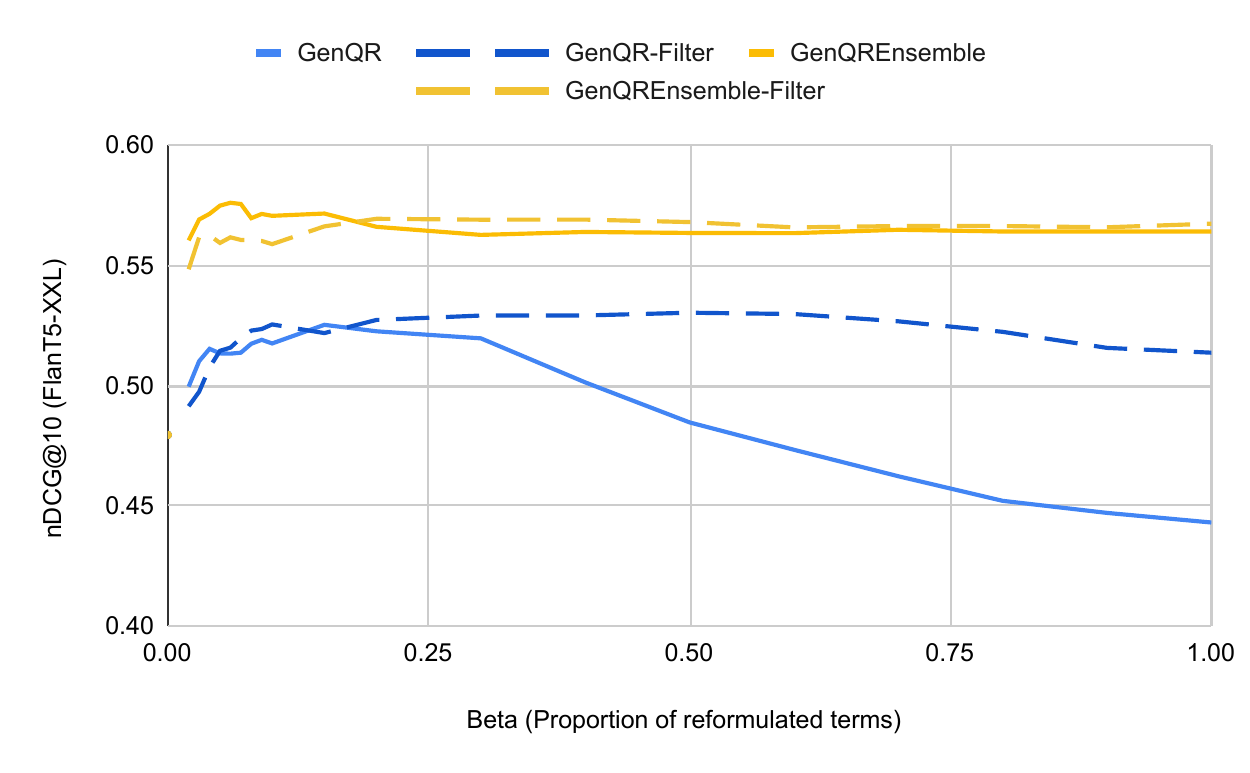}
\end{minipage}
\caption{\blt{Effect of Filtering on Llama2 (left) and FlanT5 (right) generated keywords for TP19.}}
\label{fig:filtered}
\end{figure*}

The enhanced effectiveness of the ensemble approach can be attributed to the sensitivity of LLMs to inputs with minor differences~\cite{robust1, robust2}. Investigating if the models are equally responsive to paraphrastic instructions that emphasize domain-specific keywords might be beneficial to improve performance over specific target domains. We hence measure if domain-specific instructions can further improve the retrieval effectiveness of the target benchmark. We modify the initial instruction $I_1$ to instruct the model to take into consideration the target domain. For instance, for DBPedia Entity Retrieval, we define $I_1$ as shown in Figure~\ref{fig:domainprompt}. We then paraphrase as before to generate $N=10$ instructions and re-evaluate retrieval performance.

The results are shown in Table~\ref{llama2DS}. Using domain-specific instructions improves over generic instructions across all the benchmarks in the BM25 single instruction setting. In the ensemble setting, nDCG@10 improves for the three domain-specific datasets - TR04, WT, and DB. In the neural setting, the DB Entity Retrieval sees improvement in both single instruction and ensemble settings.

\subsection{Filtering Query Reformulations}
We noticed that many of the produced reformulations are verbose, and hence we perform automated filtering to keep only the most relevant keywords. For this task, we prompt the GPT-4 API to filter the query reformulations from~\texttt{flan-t5-xxl} as well as~\texttt{llama-2-7b-chat-hf} and measure the retrieval effectiveness of the reformulations (shown as FL in Figure~\ref{ensembleperformance_bm25}). The results for different proportions of the reformulated and filtered queries are shown in Figure~\ref{fig:reformweights} and~\ref{fig:filtered}. With the addition of a filter layer, the performance almost remains the same but the interpretability of the queries increases drastically (described in detail in subsection~\ref{sec_nlr}).

\subsection{Exploring Interpretable Reformulations}
\label{sec_nlr}
\begin{table}[h]
    \centering
    \resizebox{\columnwidth}{!}{%
    \begin{tabular}{ll|c|r}
         & Comparisons & nDCG@10 & Interpretability \\ 
        \hline
        \multirow{2}{*}{GenQR} & KW vs NL & \textbf{.496} vs .479 & 36 vs \textbf{50} \\ 
        ~ & KW vs FL & \textbf{.496} vs .493 & 3 vs \textbf{83} \\ 
        \hdashline
        \multirow{2}{*}{\NameQR{}} & KW vs NL & \textbf{.580} vs .529 & 37 vs \textbf{49} \\ 
        ~ & KW vs FL & \textbf{.580} vs .575 & 5 vs \textbf{81} \\ 
    \end{tabular}}
        \caption{Comparison of Keyword reformulations (KW) with Filtered (FL) and Natural Language reformulations (NL) alongwith interpretability preferences out of 86}
    \label{interpretability}
\end{table}
While keyword-based reformulations generated from LLMs improve retrieval effectiveness on multiple benchmarks, they can sometimes be messy and hard to interpret and may lack fluency to the level desired for other downstream applications like question answering. Some studies~\cite{qw1, qw2, qw3} have emphasized the importance of fluent reformulations or generated well-formed expansions~\cite{wang-etal-2023-query2doc}. In that regard, we modify the instructions to elicit natural language reformulations (shown in Figure~\ref{fig:nl_prompt}) and attempt to generate comparatively fluent and interpretable reformulations rather than keywords. In addition to retrieval effectiveness, we perform a model-based evaluation using GPT-4~\cite{openai2023gpt4} to compare the interpretability of keyword-based queries and natural language queries through the prompt shown in Figure~\ref{fig:evaluation_prompt}. The order of the queries being compared is reversed and evaluation (on 43 queries) is reperformed to mitigate possible position bias~\cite{Wang2023LargeLM}. We find that while nDCG@10 drops slightly, reformulations are comparatively more interpretable and easy to understand. Filtered queries are also preferred against their pre-filtered counterparts. The results are shown in Table~\ref{interpretability}. 
\begin{figure*}
\centering
\fbox{
\begin{tabular}{l}
Which one of the following query reformulations for the original query ... is more 
    interpretable and \\easy to comprehend and understand for a reader. First analyze both the reformulations and provide 
    a \\short explanation why one is more interpretable than the other and then specify reformulation A or \\reformulation 
    B as your final option \\ Reformulation A: ... \\ Reformulation B: ...
    \\ Specify either A or B.
\end{tabular}}
\caption{Evaluation Prompt Used to Measure Interpretability of the Generated Reformulations. The evaluation was performed with the order of the placeholders reversed too.}
\label{fig:evaluation_prompt}
\end{figure*}
\begin{figure}[h]
\centering
\resizebox{\columnwidth}{!}{%
\fbox{
\begin{tabular}{l}
You are a helpful assistant who directly provides a \\natural language 
    reformulated query with \\novel keywords related to the user's original query. \\Do not explain yourself. Just return 
    a natural language query.
\end{tabular}}}
\caption{Prompt Used to Generate Natural Language Reformulations}
\label{fig:nl_prompt}
\end{figure}
\section{Conclusions}
Zero-shot QR is advantageous since it does not rely on any labeled relevance judgments and allows eliciting pre-trained knowledge in the form of keywords by prompting the model with the original query and appropriate instruction. By introducing \textbf{\NameQR{}} and \textbf{\NameDF{}}, we show that zero-shot performance can be further enhanced by using multiple views of the initial instruction, both as a unified query and through document fusion. We also show that the PRF extension \textbf{\NamePRF{}} can effectively incorporate relevance feedback, either automated or from users. With domain-specific instructions, we can further incorporate specific information to improve effectiveness over benchmarks with specific focus. A final filtration step to convert messy keywords to their fluent counterparts helps increase interpretability. Our proposed ensemble approach improves upon the state-of-the-art zero shot reformulation and can be applied to a variety of settings, for example, to address different aspects of queries or metrics to optimize, or to better control the generated reformulations, or for improving queries for retrieval augmented generation.

\section{Limitations}
While generative QR greatly benefits from our ensemble approach, the proposed methods come at a cost of potentially increased latency, but this is becoming less problematic with the increased availability of batch inference for LLMs. 

In our work, we have presented a zero-shot approach for incorporating domain-related information through domain descriptions. There are other ways to incorporate domain information like presenting exemplar documents of the target domain or sample terms or phrases of the same. Our objective was to establish the benefits of ensembling while also presenting baselines of ancillary avenues of increasing performance. Further performance enhancement could be achieved by resorting to each of those directions. 

The interpretability of not only query reformulations but even of other language phenomena is often highly subjective~\cite{miller2019explanation} and it could vary according to the intended application. Besides, it could be argued that natural language might not always be the best mode for interpretability. While natural language expressions could communicate the precise intent, keywords could also be useful for clustering or visualization -- and hence both being useful for interpretability. Our work closely adheres to the former definition of interpretability. 

\section{Ethical considerations}
Large Language Models should be conceptualized as socio-technical subsystems~\cite{selbst2019fairness, dhole-2023-large}. Although our research did not identify harmful outputs, it is essential to acknowledge that other instances, might produce toxic or harmful keywords. Consequently, reformulators must undergo rigorous testing before deployment to ensure that the generated content, particularly keywords, does not exhibit toxicity or bias. Such precautionary measures are critical in mitigating potential risks associated with the deployment of LLMs.

\bibliography{custom}
\end{document}